\documentclass[sigconf]{acmart}

\usepackage{enumitem}
\usepackage{comment}
\newlist{questions}{enumerate}{2}
\setlist[questions,1]{label=RQ\arabic*.,ref=RQ\arabic*}
\setlist[questions,2]{label=(\alph*),ref=\thequestionsi(\alph*)}

\usepackage{multirow}

\copyrightyear{2024} 
\acmYear{2024} 
\setcopyright{rightsretained} 
\acmConference[Websci '24]{ACM Web Science Conference}{May 21--24, 2024}{Stuttgart, Germany}
\acmBooktitle{ACM Web Science Conference (Websci '24), May 21--24, 2024, Stuttgart, Germany}\acmDOI{10.1145/3614419.3644023}
\acmISBN{979-8-4007-0334-8/24/05}

\begin{document}

\title{Shorts vs. Regular Videos on YouTube: A Comparative Analysis of User Engagement and Content Creation Trends}

\author{Caroline Violot}
\email{caroline.violot@unil.ch}
\orcid{}
\affiliation{%
  \institution{University of Lausanne}
  \streetaddress{}
  \city{}
  \country{}
}

\author{Tuğrulcan Elmas}
\email{telmas@iu.edu}
\orcid{}
\affiliation{%
  \institution{Indiana University Bloomington}
  \streetaddress{}
  \city{}
  \country{}
}
\author{Igor Bilogrevic}
\email{ibilogrevic@google.com}
\orcid{}
\affiliation{%
  \institution{Google}
  \streetaddress{}
  \city{}
  \country{}
}
\author{Mathias Humbert}
\email{mathias.humbert@unil.ch}
\orcid{}
\affiliation{%
  \institution{University of Lausanne}
  \streetaddress{}
  \city{}
  \country{}
}


\begin{abstract}
 YouTube introduced the \emph{Shorts} video format in 2021, allowing users to upload short videos that are prominently displayed on its website and app. Despite having such a large visual footprint, there are no studies to date that have looked at the impact Shorts introduction had on the production and consumption of content on YouTube. 
%
This paper presents the first comparative analysis of YouTube Shorts versus regular videos with respect to user engagement  (i.e., views, likes, and comments), content creation frequency and video categories. We collected a dataset containing information about 70k channels that posted at least one Short, and we analyzed the metadata of all the videos (9.9M Shorts and 6.9M regular videos) they uploaded between January 2021 and December 2022, spanning a two-year period including the introduction of Shorts. Our longitudinal analysis shows that content creators consistently increased the frequency of Shorts production over this period, especially for newly-created channels, which surpassed that of regular videos. We also observe that Shorts target mostly entertainment categories, while regular videos cover a wide variety of categories. In general, Shorts attract more views and likes per view than regular videos, but attract less comments per view. However, Shorts do not outperform regular videos in the education and political categories as much as they do in other categories.
Our study contributes to understanding social media dynamics, to quantifying the spread of short-form content, and to motivating future research on its impact on society.
\end{abstract}

\ccsdesc[500]{Information systems~Social networks}
\ccsdesc[500]{Information systems~Web log analysis}


\keywords{YouTube, Short-Form Video Content, Engagement, Popularity, Upload Behavior, Social Media Dynamics, Content Production Patterns}



\maketitle

\section{Introduction}

During the last few years, short-form video content has gained widespread popularity~\cite{forbesPopularityShorts,southern_youtube_2022}. TikTok, a social media platform launched in 2016 that focuses on short videos, quickly became a commercial success, with 3 billion downloads and 1 billion active monthly users in 2023 \cite{tiktokstats}. Shortly after that, platforms such as YouTube, Instagram, and Facebook introduced their own short-form video content features, with a similar format across all platforms. 
YouTube, in particular, introduced its so-called \emph{Shorts} format as a beta version in the US on March 18, 2021~\cite{ytShortsUSIntroduction}, and worldwide a few months later. 
Since its introduction on YouTube, more and more content creators have started to produce content in this format~\cite{southern_youtube_2022}. 

The world of short-form video content is currently a topic of lively discussion, with a spectrum of perspectives. On the one hand, it provides an opportunity for creators to engage and entertain their audience with concise and attractive content. Moreover, short-form videos can be effective at disseminating information about social issues~\cite{9758751} and create new professional perspectives for content creators~\cite{9211390}. On the other hand, recent studies exploring TikTok highlighted some potentially concerning aspects for its users, such as a form of dependency~\cite{su2021viewing}, increase of daytime fatigue~\cite{Wang_Scherr_2022} and decrease of prospective memory~\cite{Chiossi_Haliburton_Ou_Butz_Schmidt_2023}. Moreover, its impact on informative content, for example educational videos that thrive on depth and detail, should be explored. 
To investigate this, we focus on a fundamental research question: “Are short videos replacing longer videos on YouTube, the most popular online video-sharing platform?"

To address this question, we undertake the first comprehensive study of YouTube Shorts, comparing them with regular videos (RVs), in terms of their effect on overall channel behavior and user engagement. Using data from the public YouTube Data API~\cite{youtubedataAPI}, we are able to control for platform, creator, and video features, offering insights into the evolving YouTube ecosystem. Our longitudinal comparative analysis quantifies changes in video content creation and user engagement across Shorts and RVs over the two years following the introduction of YouTube Shorts in March 2021. This study aims to provide valuable insights to content creators, advertisers, and researchers, allowing them to better understand the perspectives of an era marked by the rise of short-form video content. To delve deeper into the impact of Shorts on platform content and user engagement, we address the following research questions:
\begin{enumerate}[leftmargin=1.5\parindent,align=left,labelwidth=\parindent]
\item[RQ1] How did the introduction of Shorts affect preexisting channels in terms of content creation behavior, and how do channels created after Shorts introduction differ from older channels?\label{RQ:1}
\item[RQ2] How do Shorts compare to RVs in terms of views and content creation frequency across video categories? \label{RQ:2}
\item[RQ3] What differences between Shorts and RVs can be observed in terms of user engagement (views, likes, and comments), and does the duration of RVs have an influence?\label{RQ:3}
\end{enumerate}

Our dataset contains data about 70k channels that have created at least one video in the Shorts format since March 2021. We collected metadata of 9.9M Shorts and 6.9M RVs posted by those channels between January 2021 and December 2022, which allowed us to analyze both the time window around Shorts introduction and the long-lasting impact it had on those channels and their videos. 
For each video we also retrieved its category and its number of views, likes, and comments. 

Our results highlight the emergence of three main trends. First, we observed that channels that posted at least one Shorts tended to adopt the format, eventually uploading more Shorts than RVs. Second, we found that categories are not distributed evenly between Shorts and RVs: Shorts were mainly uploaded in entertainment-related categories while RVs encompassed a wide variety of content, including political or educational. This indicates that the two types of videos are not created to cover the same themes, but rather co-exist on the platform for different purposes. Additionally, political, educational and artistic Shorts videos generate fewer views, suggesting that for some categories, viewers prefer RVs. 
Overall, we found that Shorts outperformed RVs in terms of views and likes per view, but generated less comments per view, although this gap is narrowing. This trend is even more pronounced when comparing the engagement metrics of Shorts and RVs uploaded by the same channel, with Shorts getting 110 times more views (on average) than their RVs counterparts. However, when differentiating RVs between different duration groups we found that the median number of views of videos from 10 to 30 minutes long was higher than the median number of Shorts views. 

The rest of the paper is organized as follows. Section \ref{sec:data_collection} describes how we collected and processed the data used in this article. Section \ref{sec:posting_dynamics} shows how the video publishing behavior evolved over the period of observation. Section \ref{sec:content_differences} reports the differences in content between Shorts and RVs. Section \ref{sec:engagement} describes how users engaged with both types of videos in terms of views, likes, and comments. Section~\ref{sec:discussion} discusses the impact of Shorts' introduction on overall content publishing behavior and user reaction.
Section~\ref{sec:related_work} provides the prior literature on the topic. Finally, Section~\ref{sec:conclusion} concludes the paper with its main findings, discusses its limitations, and provides future directions of research.

\section{Data Collection}\label{sec:data_collection}

\begin{figure*}    \includegraphics[width=\textwidth]{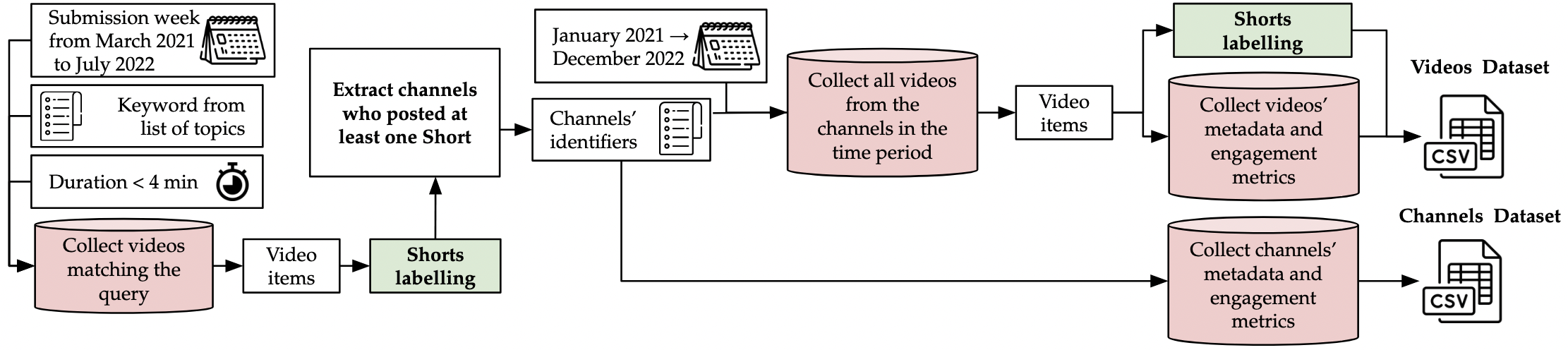}
\caption{Data collection process summary. }
\end{figure*}

Before diving into the details of our data collection, hereafter we briefly describe YouTube's RVs and Shorts.
RVs can last from a few seconds to several hours. They are recorded and usually edited outside of YouTube, and they can be in a horizontal, vertical, or square format \cite{uploadvideos}. Shorts is a newer format that can last up to 60 seconds and must be in a vertical or square shape, optimized for viewing on mobile devices. 
They can be created outside of YouTube and then posted, but they can also be created directly from the app, by filming one or several clips that are combined on the spot, adding music, adjusting recording speed, adding filters, etc. \cite{createShorts}, making Shorts particularly easy and quick to shoot, edit, and upload. 
Shorts have their own dedicated tab on the platform website or app, and users can move from one video to another by swiping on a endless scroll, without actively clicking on or searching for the videos.

In order to efficiently collect relevant data, we leverage the YouTube Data API \cite{youtubedataAPI}. It provides methods to search video metadata by keywords or by channel identifiers, and to collect channel metadata, among other functionalities. Unfortunately, the YouTube Data API does not currently support random sampling of videos, hence we had to define a methodology to collect videos while acknowledging that the resulting dataset could contain some biases. We discuss them at the end of this section.

%

\subsection{Collection Process}

The data collection consisted of three steps: (i) collecting an initial set of short videos (i.e., seeds), (ii) identifying which of them are Shorts, and (iii) growing the dataset.\\

\subsubsection{Collecting seed Shorts}
Our primary objective is to collect seed Shorts so that we can identify channels that include both Shorts and RVs for our comparative analysis. As the YouTube API requires keywords to provide videos, we first come up with a comprehensive set of keywords that represent video categories for which the YouTubers create Shorts. To this end, we used common video categories of TikTok as search queries, assuming they would also be common in YouTube Shorts. We collected the categories from a digital marketing website.\footnote{\url{https://marketsplash.com/tiktok-hashtags/\#link3}} We separated the terms that contained an ``\&'' (for example ``food \& cooking'' was split into a ``food'' keyword and a ``cooking'' keyword), for a total of 50 keywords. 

The queries returned both Shorts and RVs as the YouTube API did not provide an option to collect only Shorts. However, it was possible to restrain the ``Search'' results to ``short'' (less than 4 minutes), ``medium'' (between 4 and 20 minutes), and ``long'' videos (more than 20 minutes). Thus, to maximize the amount of Shorts we collected in this seed phase, we collected only videos from the ``short'' category.


We first collected videos posted between March 18, 2021, (the date of the US beta launch of YouTube's Shorts \cite{ytShortsUSIntroduction}) and July 26, 2022. The API returns around 500 videos per query, so to further increase the number of results and ensure that the search results were not biased towards a specific period, we divided the time period into weeks, e.g., we first collected videos that were published between March 18, 2021, and March 21, 2021,\footnote{First week is incomplete as March 18, 2021 was a Thursday.} then collected videos between March 22, 2021, and March 28, 2021, and so on, each time using all the aforementioned keywords. In total, we collected around 300k videos from 150k channels.\\

\subsubsection{Labelling Shorts}
The YouTube Data API does not currently return information on whether a video is a Short or a RV. Hence, we use the following methodology to identify which videos are Shorts: we send a GET request to \path{www.youtube.com/shorts/<videoId>} for each videoId and check in the redirection link if the URL stayed the same or if it was modified to the regular \path{www.youtube.com/watch?v=<videoId>}. 
YouTube allows older videos to be seen in the Shorts tab, as long as they are up to 60 seconds and have a square or vertical aspect ratio, so this method can classify videos uploaded before Shorts introduction as Shorts. Nevertheless, this method allows us to determine whether creators turned to short-form, square/vertical video content. We classified 144k videos as Shorts, and 159k as RVs using this method.\\

\subsubsection{Growing the dataset and collecting additional metadata}
Using the aforementioned method to identify Shorts, we identified all the channels that contained at least one Short during our period of interest. Among the 150k channels collected in the first step, there were 70,712 channels with at least one Short video. We collected all the videos posted by these channels between January 1, 2021, and December 31, 2022, totalling 16,746,091 videos, among which 6,862,321 RVs and 9,883,770 Shorts.
Using the YouTube Data API, we collected the videos' metadata (title, description, posting date and time, duration, channel) and engagement statistics (number of views, likes, and comments). We also collected the YouTube categories of each video. 
YouTube categories (listed in Table \ref{tab:YT_cat} in Section \ref{sec:content_differences}) consists of 15 categories that creators or YouTube assign such as \textit{Music} or \textit{Gaming}. 
Finally, we collected the channels' metadata, mainly their title, description, creation date, and origin country. We further collected the channels' engagement metrics, i.e., the total view count, subscriber count, and number of uploaded videos.




\subsection{YouTube Terms \& Conditions Compliance}

The YouTube Data API limits the daily number of queries with a quotas system, where costs vary between different methods, e.g., a video search query costs 100 quotas and a channel metadata query costs 1 quota. The default number of quotas per day is 10k. As this limit is too restrictive to collect a large-scale dataset, we applied to and joined the YouTube Research Program~\cite{youtubeResearchProgram} and obtained a research quota extension of 1M queries per day. We made sure to comply with the specific data policies and terms of use that come with being part of the YouTube Research Program \cite{youtubeResearchTerms}.
In particular, we are not able to share our data due to the no-data disclosure, which forbids us to ``disclose, reproduce, sell, license or otherwise transfer to any third party, in part or in whole, any Program Data''. 

\subsection{Bias Discussion}

Ideally, we would prefer a random sample of videos for our collection of seed Shorts, for an unbiased analysis. However, as previously mentioned, YouTube does not currently support such random sampling. Furthermore, we are not able to use an empty query or queries with very general keywords to approximate random sampling. This is because we observe that regardless of the keyword, YouTube limits the search results amount, e.g., we could only collect 597 videos with the query ``cats'' and 251 videos with an empty query. Past research instead focused on videos that were the most influential using popularity as a proxy. For instance, Riberio et al. crawled channels with at least 10k subscribers and then collected their videos to provide a comprehensive dataset of YouTube~\cite{ribeiro2021youniverse}. Although this approach may facilitate studying popular channels, it may prevent us from analyzing YouTube Shorts that went viral and were viewed many times despite the low popularity of the channel. Our approach mitigates such a bias, but we acknowledge that it creates a bias towards the videos related to the keywords we used. 


\section{Posting Behavior Evolution}\label{sec:posting_dynamics}

We present here the evolution of the channels posting behavior, from January 2021 to December 2022. 

\subsection{Evolution of Global Video Uploads}

\begin{figure}
    \centering \includegraphics{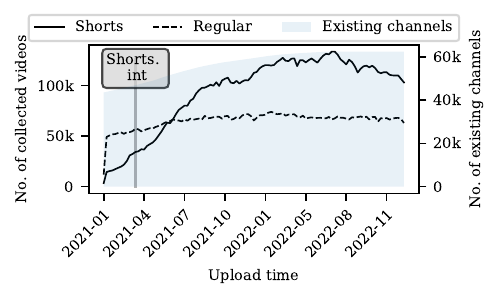}
    \caption{Weekly video uploads, categorized into Shorts and RVs, with the number of channels older than the respective week shown in light grey. Some videos created before the introduction of Shorts were retrospectively classified as Shorts. 
    }
    \label{fig:global_weekly_uploads}
\end{figure}

We first focus on the overall posting evolution, without considering individual channels behavior and analyze the total number of Shorts and RVs uploaded per week. YouTube allows any videos shorter than 60 seconds, with a square/vertical format, to be displayed in the Shorts tab (and therefore categorized as such), hence videos from before Shorts introduction can be labeled as such. As shown in Figure \ref{fig:global_weekly_uploads}, we observe a constant rise in the number of new Shorts until mid-2022, followed by a slight decrease, and a constant number of created RVs. This shows that, while collectively continuing to produce RVs, video creators have also produced an increasing number of Shorts since March 2021. 

\subsection{Evolution of Shorts Prevalence}
Next, we focus on the individual posting behavior of channels, aiming to show results that equally reflect the behavior of all the channels in our dataset, without being biased towards the highly prolific channels. For most of the analysis, we first split the channels between \textit{newer channels} and \textit{older channels}, based on whether they were created after or before Shorts introduction.

\begin{figure}
    \centering    \includegraphics{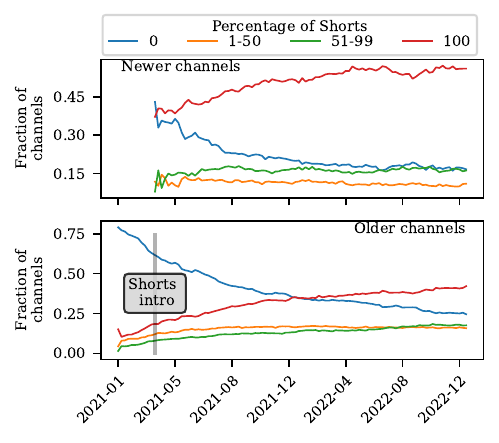}
    \caption{Analysis of channels' posting activity from January 2021 to December 2022. Channels were divided into groups based on the percentage of Shorts in the videos uploaded each week. The evolution of the fractions of channels in each group is shown. 
    }
    \label{fig:channels_cat_Shorts_percent}
\end{figure}

For each week between January 2021 and December 2022, we categorize active channels based on the percentage of Shorts they posted each week into the following categories: [0\%, 1-50\%, 51-99\%, 100\%]. A given channel can change category from one week to another, if its posting behavior evolves. Then, we compute the fraction of channels belonging to each category and show the evolution of the posting behavior in Figure \ref{fig:channels_cat_Shorts_percent}. Channels are separated between older channels and newer channels. 

In March 2021, the fraction of channels posting exclusively Shorts was 2.2 times higher among newly created channels than for older channels, with more than 60\% of the latter opting to posting exclusively RVs. From there, we see a similar evolution for older and newer channels, with a marked increase in the fraction of channels posting only Shorts and a decrease in the fraction of the channels posting only RVs. However, while the intermediate categories exhibit a constant fraction for newer channels, we observe that, for older channels, the intermediate categories increased since January 2021. This indicates that a substantial number of the older channels which started posting Shorts continued to create RVs as well. 

\subsection{Evolution of Posting Frequency} 

\begin{figure}
    \centering    \includegraphics[width=\columnwidth]{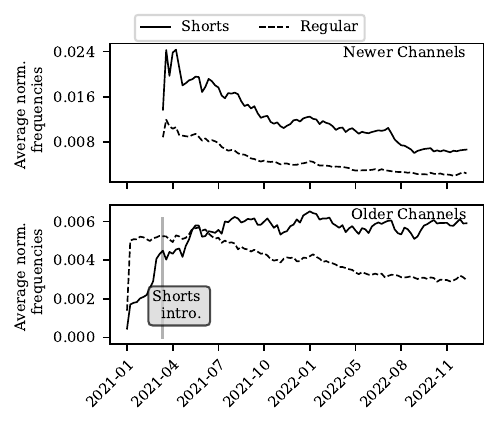}
    \caption{Evolution of normalized uploads frequency. Average weekly uploads of Shorts and RVs of each channel normalized by the total number videos (of both types) posted by that channel. }
    \label{fig:posting_frequency_evolution}
\end{figure}

Having observed that channels created an increasing number of Shorts, collectively and individually, we analyze the impact it had on the production of RVs. 
Again, separating channels between older and newer channels, for each channel $c$ and each week $w$, we compute $n_{cw}^{regular}$ (resp. $n_{cw}^{Shorts}$), the number of RVs (resp. Shorts) uploaded that week by that channel. We divide each $n_{cw}^*$ by $n_c$, the total number of videos posted by channel $c$, and obtain $f_{cw}^*$, the normalized frequency of posting for each channel. We finally compute $f_w^*$,  by averaging the normalized frequencies across all channels for each week, shown in Figure~\ref{fig:posting_frequency_evolution}. This approach enables us to discern the weeks during which channels collectively uploaded more or fewer videos of each type, each channel contributing equally to the outcome, regardless of their total uploads count. 

We observe that both newer and older channels progressively reduced the frequency of RVs uploads over time.  However, while the frequency of Shorts uploads increased for older channels, it decreased for newer channels. This is surprising at first, considering that some of the channels were created after and therefore their highest posting frequency should increase the value of the average frequency on later weeks. One possible explanation is that many channels which appeared around Shorts introduction were created in order to try the Shorts format, but lots of them stopped posting videos after a few weeks (out of the 445 channels created between 18 March 2021 and 25 March 2021, 4\% had posted half of their videos a week after their creation and 9\% had posted half of their videos a month after their creation) and therefore have a high normalized frequency of posting during their first few weeks of activity.

\subsection{Evolution of Weekly Content Volume}

\begin{figure}
    \centering
    \includegraphics{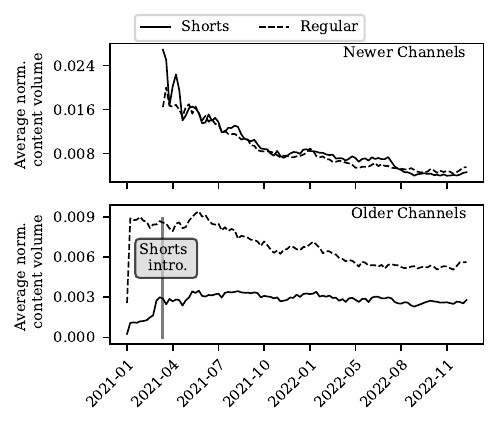}
    \caption{Evolution of normalized weekly content volume.  Average over all channels of the weekly content volume (sum of the durations of each video posted that week), separated between Shorts and RVs, normalized by the combined weekly content volume of both types.}
    \label{fig:content_volume}
\end{figure}

To complement the evolution of content uploading, we also analyzed the sum of the durations of the videos uploaded each week, called the weekly \textit{content volume}. 
This allows us to get a sense of the amount of Shorts and RV content produced each week. As before, we separate our results between older and newer channels. 
Following the same logic as for the normalized frequency of posting, for each channel $c$ and each week $w$, we compute $d_{cw}^{regular}$ and $d_{cw}^{Shorts}$, respectively the RVs' content volume and the Shorts' content volume uploaded that week by that channel. 
For~$* \in \{regular,\: Shorts\}$, we divide each $d_{cw}^*$ by $d_c$, the total content volume of channel $c$, and obtain $v_{cw}^*$, the normalized weekly content volume for each channel. 
The resulting quantities, shown in Figure~\ref{fig:content_volume}, allow us to see on which week did channels invest the more time on average. 
Similarly to posting frequencies, newer channels had a peak of content creation around the introduction of Shorts and rapidly decreased from there, for both Shorts and RVs. 

As for older channels, since the introduction of Shorts, on average channels have increased their amount of Shorts content until reaching a plateau around June 2021, but the amount of RVs content is declining. 

\section{Content Analysis}\label{sec:content_differences}

Our content analysis relies on video categories. The category is unique to a video and is either chosen by the creator or assigned by YouTube. Public videos can be assigned 15 categories, the others being movie genres for paid content.

We first examine the distribution of categories for Shorts and RVs. Table \ref{tab:YT_cat} shows the categories, and the number of videos we collected for each category. \textit{People \& Blogs} being the default category it is over-represented. In the second place, we find the \textit{Entertainment} category. \textit{Nonprofits \& Activism} and \textit{Pets \& Animals} were the categories for which we collected the fewest videos. 

In general, the fraction of Shorts differs widely between categories. 
Categories with the highest fractions of Shorts are \textit{Comedy} and \textit{People \& Blogs} whereas \textit{News \& Politics} and \textit{Nonprofits \& Activism} exhibit the lowest fractions. 
This observation may suggest that Shorts are predominantly used for generating lighthearted content, while RVs are the preferred format for delivering more serious information. \textit{People \& Blogs} having the largest number of Shorts could imply that creators do not specify the category when posting Shorts as often as they do when posting RVs, or that YouTube takes longer to classify them into the relevant categories. 

\begin{table}
\small
\addtolength{\tabcolsep}{-2.5pt}
    \centering
    \begin{tabular}{c||c|c}
        \multirow{2}{*}{YouTube categories} &  Collected &\multirow{2}{*}{\% of Shorts} \\
         & videos count & \\
         \hline
         \hline
People \& Blogs & 5.7M & 74.3\\
Entertainment & 3.2M & 55.5\\
Howto \& Style & 1.9M & 57.3\\
Education & 1.8M & 43.8\\
Gaming & 1.0M & 54.4\\
News \& Politics & 766.4k & 14.4\\
Sports & 544.8k & 39.3\\
Comedy & 385.3k & 76.5\\
Science \&
 Technology & 340.3k & 51.2\\
Music & 333.0k & 63.2\\
Film \& Animation & 288.9k & 57.3\\
Travel \& Events & 208.0k & 57.3\\
Autos \& Vehicles & 200.3k & 57.8\\
Pets \& Animals & 118.4k & 68.5\\
Nonprofits \& Activism & 42.5k & 34.1\\
    \end{tabular}
\caption{YouTube categories and the corresponding number of videos from our dataset. The percentage of Shorts out of all the videos collected in each category is also shown.}
    \label{tab:YT_cat}
\end{table}

\begin{figure*}
    \centering
    \includegraphics{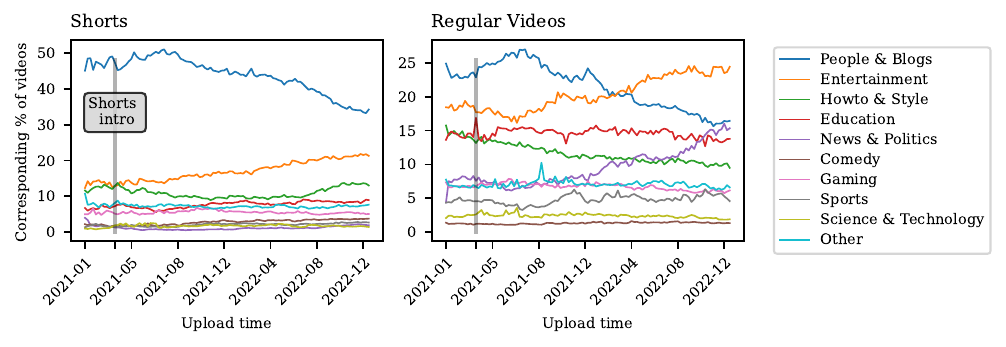}
    \caption{Evolution of the percentage of categories attributed to Shorts and RVs, showing the changes of categories popularity among creators over our time period.}
    \label{fig:cat_evol}
\end{figure*}

We then observe the evolution of the categories over our two years period. We selected the nine most common categories attributed to the videos in our dataset and labelled the rest of them under ``Other''. Next, for each category, we compute the percentage of videos to which the category was attributed out of all the videos posted in a given week, and repeat that for every week of our time period. Results are shown in Figure \ref{fig:cat_evol}. We see that Shorts videos are consistently dominated by the \textit{People \& Blogs} category, although it is slowly declining in favor of the \textit{Entertainment} category. 

Categories of RVs are way more diverse and evenly distributed, and we can observe large trends of categories rising, declining or maintaining a constant percentage.  The \textit{People \& Blogs} and \textit{Entertainment} categories are also the most commonly attributed, but not as dramatically above as for Shorts. We notice that the \textit{Education} category and the \textit{Science \& Technology} category maintained a constant percentage of uploads, and that the \textit{News \& Politics} category steadily increased after the beginning of 2022.

\section{Engagement Analysis}\label{sec:engagement}

In this section, we examine the evolution of the number of views, likes and comments, collectively referred to as the engagement metrics, received by Shorts and RVs. Our caveat is that we only have access to the engagement metrics as of the query time, which limits the analysis of the metrics evolution. However, most videos experience their peak of attention a few days after their release \cite{cha_analyzing_2009}, nine months passed between the latest video's publication date (December 31, 2022) and the collection of engagement statistics (September 1, 2023). Therefore, besides a few exceptions where videos become viral a long time after their publication date, engagement metrics should not drastically fluctuate, and rather grow at a seemingly constant rate. 
This allows us to draw comparative results between the popularity of Shorts and RVs.

\subsection{Engagement at the Video Level}

We first analyzed the engagement at the video level, aggregating the results without considering channels or categories. Coherently with previous work \cite{Arthurs_Drakopoulou_Gandini_2018}, we found that 1\% of the Shorts (resp. RVs) attracted 63\% (resp. 61\%) of the Shorts (resp. RVs) views. We also computed that views and likes are highly correlated, with a Pearson correlation coefficient (PCC) of 0.848, but that the number of comments is not necessarily correlated to the others (with a PCC of 0.273 with views and a PCC of 0.360 with likes). 

\begin{figure}
    \centering
    \includegraphics{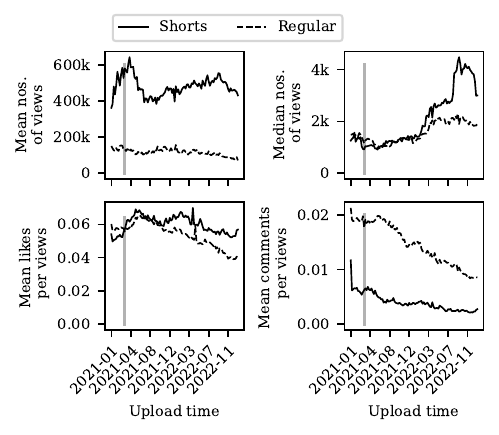}
    \caption{
  Evolution of engagement metrics for Shorts and RVs, including the mean views, median views, mean of the likes count divided by views count and mean of the comments count divided by views count. The gray line indicates Shorts introduction.}
    \label{fig:engagement_metrics}
\end{figure}

In Figure \ref{fig:engagement_metrics}, we present an overview of the engagement metrics evolution. Specifically, we tracked the mean number of views, the median number of views, the mean likes per view and the mean comments per view.


We first observe that, over the two years, Shorts received four times as many views as RVs, and by the end of 2022, this difference had increased to six times. This is not surprising given that each user may watch far more Shorts than long videos in the same amount of time. We also note that mean views of RVs declined slowly and consistently with time, whereas mean views of Shorts fluctuated around its introduction and then mostly increased from there, before slightly declining near the end of 2022. 

As previously mentioned, the vast majority of views are harvested by a handful of videos. The extreme engagement values obtained by the top 1\% of videos skew the means towards higher values and prevent from grasping the engagement evolution of more low-ranking videos which constitute the majority of videos. Looking at the median allows to better understand the dynamics of the 99\% majority. Since the introduction of Shorts in March 2021 and until the end of the year, we observe a consistent and similar increase in the median number of views, with a slope of increase of 11.9 ($r^2 = 0.74$) for RVs and a slope of increase of 13.5 ($r^2: 0.81$) for Shorts. But while RVs reached a plateau around March 2022, the median number of views that Shorts attracted increased drastically during 2022, with a slope of increase of 53.0 ($r^2 = 0.78$).
This shows that even less popular Shorts still obtain a substantial number of views which is not the case of RVs. The Shorts format would then allow not yet popular creators to reach a wider audience than RVs.


Regarding the other engagement metrics, we see that around Shorts introduction, Shorts and RVs have the same likes per view rate, but as from August 2021, Shorts started to convert views into likes more effectively than RVs and by the end of 2022, Shorts' likes per view rate was 1.4 times higher than RVs' likes per view rate. Comments per view, a more active form of engagement \cite{cha_analyzing_2009}, have a higher rate for RVs than Shorts, but the gap seems to be narrowing over time. 

\subsection{Engagement at the Channel Level}

We established that Shorts are generally more viewed and liked than RVs. We now explore if that is the case for videos originating from the same channel and if the trends that were observed globally also apply on a channel basis. 
We first split our channels between the 1\% with the most subscribers (referred to as \emph{top 1} channels) and the rest (referred to as \emph{bottom 99} channels). The top 1\% channels views accounted for 46\% of the total views.  

\begin{figure}
    \centering
    \includegraphics[width=\columnwidth]{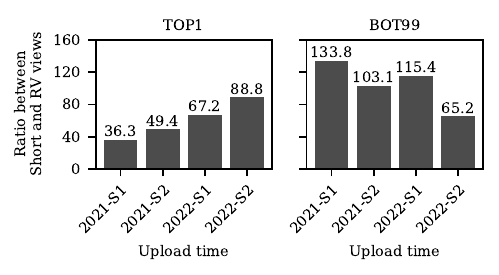}
    \caption{Evolution of the average ratio between channels' Shorts' and RVs' views, divided between the top 1\% of channels with regards to subscribers count and the rest.}
    \label{fig:channels_engagementl}
\end{figure}

We divide our two-year period into four semesters referred to as ``2021-S1'' for the 1$^{\text{st}}$ semester of 2021, ``2021-S2'' for the 2$^{\text{nd}}$ semester of 2021, and so on. We then compute the mean views per channel and per semester, distinguishing between Shorts and RVs. This process is repeated for each semester, including only channels that posted both Shorts and RVs that semester. The ratio of mean views for Shorts to RVs is computed for each channel, and the average ratio is obtained by averaging across all eligible channels for each semester. This yields the evolution of the average ratio between Shorts' and RVs' views, on a channel basis, and a closer intuition to the difference  between Shorts and RVs engagement that creators can expect for their channel. Results are shown in Figure \ref{fig:channels_engagementl}.

For both popularity classes, Shorts consistently draw significantly more views than RVs from the same channel --80 times more for top 1 channels and 111 times more for bottom 99 channels, on average over the two-year period. However, while the ratio between Shorts' and RVs' views increases for top 1 channels, it declines for bottom 99 channels. Nonetheless, both groups can still reach a broader audience using Shorts than RVs.

\subsection{Engagement Based on Duration}

\begin{figure}
    \centering
    \includegraphics[width=\columnwidth]{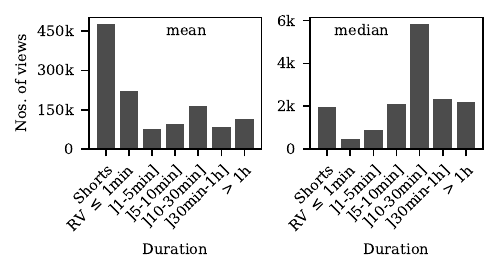}
    \caption{Mean number of views and median number of views for Shorts and RVs of different durations.}
    \label{fig:duration_engagement}
\end{figure}

We previously compared the mean and median views of Shorts and RVs but, while Shorts are restricted to 60 seconds, RVs display a wide range of durations, each format requiring a different engagement from viewers and creators. We classified the RVs based on their duration into the following time intervals: less than 1 minute, 1 to 5 minutes, 5 to 10 minutes, 10 to 30 minutes, 30 minutes to 1 hour, and longer than 1 hour. For each group we computed the mean and the median number of views. Results are shown in Figure \ref{fig:duration_engagement}.
The means of the views confirm previous results on the superiority of Shorts on RVs in terms of engagement, although at various degree, the least popular group being the 1-5 minutes group and the most popular being the under 1 minute group, followed by the 10-30 minutes group. The popularity of the later is confirmed when looking at the medians, where it appears that half of the videos between 10-30 minutes obtained around 5'800 views or more which is way above the other groups and three times as much as the median views for Shorts. This is quite surprising to observe that, median-wise, longer RVs are surpassing Shorts in attracting views.

\subsection{Engagement Based on Categories}

Finally, we compare the levels of engagement generated by Shorts and RVs within the different categories, in order to see if, for some categories, RVs attracted more views than Shorts. For each semester and each category we compute the ratio between the mean number of views per Short and the mean number of views per RV. We also compute this ratio for all categories combined, as a reference. Results are given in Table \ref{tab:categories_engagement}. 

\begin{table}[]
    \centering

\begin{tabular}{lrrrr}
\toprule
 & \multicolumn{2}{c}{2021} &\multicolumn{2}{c}{2022}\\
category    &  S1 &  S2 &  S1 &  S2 \\
\midrule
Autos \& Vehicles      &     1.42           &     2.27          &     2.49          &     2.20 \\
Comedy                 &     \textbf{7.28}  &     \textbf{4.61} &     2.77          &     3.84 \\
Education              &     \textbf{5.28}  &     3.28          &     3.88          &     4.26 \\
Entertainment          &     \textbf{6.39}  &     \textbf{4.33} &     \textbf{4.53} &     \textbf{5.70} \\
Film \& Animation      &     1.53           &     1.20          &     1.88          &     2.41 \\
Gaming                 &     2.16           &     2.45          &     2.38          &     3.97 \\
Howto \& Style         &     2.93           &     \textbf{3.55} &     \textbf{4.23} &     4.68 \\
Music                  &     0.42           &     0.46          &     0.79          &     1.10 \\
News \& Politics       &     0.72           &     \textbf{4.63} &     3.20          &     2.80 \\
Nonprofits \& Activism &     1.18           &     \textbf{4.39} &     \textbf{8.59} &    \textbf{12.88} \\
People \& Blogs        &     \textbf{5.64}  &     \textbf{4.32} &     \textbf{5.59} &     \textbf{7.05} \\
Pets \& Animals        &     2.47           &     2.84          &     \textbf{4.73} &     \textbf{8.16} \\
Science \& Technology  &     2.30           &     2.37          &     3.20          &     4.15 \\
Sports                 &     3.55           &     \textbf{3.46} &     \textbf{5.96} &     \textbf{6.59} \\
Travel \& Events       &     2.51           &     2.31          &     \textbf{4.30} &     \textbf{7.03} \\
\cmidrule(lr){1-5}
All                    &     4.19           &     3.40          &     4.04          &     5.2 \\
\bottomrule
\end{tabular}
    \caption{Evolution of the ratio between the mean number of views per Short and the mean number of views per RV, categorized by different YouTube categories, over the two years' semesters. The overall ratio trend for all categories combined is given. Values exceeding the overall ratio for each semester and category are highlighted in bold.}
    \label{tab:categories_engagement}
\end{table}

We observe that the popularity of Shorts varies between categories and time periods. It appears that for some categories, such as \textit{Music}, \textit{Film \& Animation}, \textit{Gaming}, and, to a lesser extent, \textit{Science \& Technology}, users prefer to watch RVs than Shorts. Some categories, such as \textit{Comedy} and \textit{Education}, initially generated a high engagement for Shorts before slowly loosing this advantage at the relative benefit of RVs. Conversely, for \textit{Nonprofits \& Activism}, which initially generated a limited engagement for Shorts, there was a progressive increase, eventually reaching 12 times more views for Shorts than RVs by the second semester of 2022. Finally, the \textit{Entertainment} and the \textit{People \& Blogs} categories exhibit a ratio systematically high and above the reference ratio.

\section{Discussion}\label{sec:discussion}

Our analysis sheds light on the significant impact of Shorts on the content created and consumed on the YouTube platform. 

\paragraph{RQ 1. }We focus on channels that have uploaded at least one Short to see how trying this new content affected their overall behavior. For these channels, Shorts production has grown impressively, and eventually surpassed RVs production. A notable trend is that a large proportion of channels created after March 2021 posted only Shorts from the beginning, implying that new channels were mostly created with the goal of uploading Shorts. Older channels, while initially more inclined to upload only RVs, also gradually turned to Shorts. Both older and newer channels reduced their production of RVs, and older channels persistently maintained high uploads of Shorts, further indicating that Shorts are well implanted in the YouTube landscape, and that channels have a confirmed interest in uploading Shorts, beyond the first curiosity. This growth in Shorts production may also be caused by the platform's efforts to popularize Shorts, for example by providing incentives for creators, with a new type of monetization \cite{shortsfund}, and frequent updates \cite{shorts_updates}. 


\paragraph{RQ 2.} Regarding the types of content produced, our analysis revealed that the distributions of Shorts and RVs vary widely across YouTube categories.  Shorts are primarily employed for creating entertaining content, while RVs remain the preferred format for conveying more serious information, for example on politics or social activism issues. Furthermore, we observe a synchronicity between creators and viewers, as the same entertainment-related categories exhibit a ratio of Shorts views to RV views consistently higher than the reference ratio which includes all the categories. Viewers are mainly consuming Shorts for entertainment purposes and creators may have understood that from the beginning. 

On the other hand, education-related ratio of Shorts views to RV views remains consistently below the reference ratio, indicating that, while users consume Shorts in this category, they stayed faithful to RVs for learning new things. Similarly, the exceptionally low ratio of the art-related categories suggests that users are willing to consume longer forms of content when it is videos in which artists invested time and energy. One surprising exception was the impressive engagement generated by Shorts in the \textit{Nonprofits \& Activism} category, despite the low percentage of Shorts in this category. Creators uploading videos in this category might benefit from using Shorts to reach a wider audience about social issues.

\paragraph{RQ 3.} Analyzing engagement metrics showed that Shorts are particularly effective at capturing the attention and engagement of viewers. This advantage is even more striking when we consider videos from the same channel, in which Shorts attract 110 times more views than their RV counterparts. Moreover, the gap in views between Shorts and RVs is progressively growing, both due to Shorts being increasingly watched and RVs' views declining. One nuance to this observation is that there might be a different delay between Shorts and RVs upload and their consumption by viewers. A two-year period separates the first and last videos of our dataset. Usually, after the peak of attention following their upload, videos continue to slowly accumulate views~\cite{cha_analyzing_2009}. This would partly explains why RVs posted during December 2022 have fewer views than videos from January 2021. However, newer Shorts have as many views as older Shorts, suggesting that old Shorts are rarely shown to users. 

RVs are currently having a higher comments per view rate than Shorts. Users watching RVs spend more time on the same video allowing them to engage more into comments than viewers of Shorts, rapidly swiping to the next video. Additionally, some categories prone to generate more debate and comments, like \textit{News \& Politics}, are less covered by Shorts.


On YouTube, a few videos collect the vast majority of views, and mean values are not representative of the majority of videos. Hence, to explore whether less popular channels also benefit from Shorts, we examine the median views evolution. We observe an increase of the median number of views for Shorts but also for RVs, which is contrary to the mean evolution of RVs views. The majority of Shorts are increasingly viewed but so are the majority of RVs. 

The evolution of the ratio between Shorts and RV views from the same channel confirms this particular dynamic. 
For exceptionally popular channels, Shorts are increasingly getting more views than RVs but for the remaining majority of channels the trend is the opposite.
Different reasons could explain this observation. 
The first one comes from the decreasing frequency of RV uploads. 
As most channels tend to upload fewer RVs, those that are uploaded attract more views, benefiting from some scarcity effect. 
The second one relates to the successive updates of YouTube Shorts delivery system. 
Initially accessible only from a tab on the YouTube home page, without the user being able to choose, Shorts are now recommended on the home page with thumbnails, or suggested in the ``watch next'' banner of RVs. 
These different entry points allow users to click on Shorts they want to watch out of a few suggestions, which might favor popular Shorts.

We will end this discussion with an additional nuance on the Shorts superiority over RVs. In most of our analysis we made the binary distinction between Shorts and RVs. However, RVs can widely vary in duration, and videos of 5-10 minutes constitute a type of content arguably very different than 1 hour long video essays. On average, Shorts dominate RVs of all duration groups, but at different degrees. Moreover, the analysis of the medians challenges Shorts superiority. Indeed, half of the videos in the 10-30 minutes group garnered at least 5,800 views, contrasting with a median of 1,986 views for Shorts. This duration range also ranks as the third most viewed, with a mean number of views surpassing 160k, confirming its sustained popularity among YouTube video consumers.

\paragraph{Ethical Considerations.}

As our research solely relies on publicly available data and does not involve interactions with human participants, it does not classify as human subjects research. We follow common
ethical standards; we do not attempt to de-anonymize users, we do not disclose any personal information, and our study does not report any offensive content. 


\section{Related Work}\label{sec:related_work}


Our research delves into the dynamic shifts in content creation and user behavior resulting from evolving platform policies, with a special focus on YouTube Shorts, which is a popular instance of short-form video concept. We now provide a brief survey on existing research related to short-form videos and the effect of platform policies on user behavior. 

\subsection{Short-Form Videos}

YouTube is the primary video-oriented social media platform. Past studies on YouTube focused on its role on propagating disinformation~\cite{hussain2018analyzing, hussein_measuring_2020, papadamou_it_2022}, on video popularity~\cite{borghol_characterizing_2011, wattenhofer_youtube_2012}, on attention dynamics~\cite{lee_whose_2022}, and on user interactions~\cite{wu2021crosspartisan}. Nonetheless, it is unclear if the results of these analyses focusing on regular video content hold for short-form videos as the way people interact with these two types of videos are different. 

Short-form videos are currently understudied and large-scale analyses of such data are relatively uncommon. This is first because the short-form video concept has gained widespread popularity and became the primary feature of social media platforms only recently, e.g., YouTube introduced Shorts and their section in 2021. Second, the platforms have been restrictive in their data-sharing policies. YouTube announced its research API in 2022, TikTok announced it in 2023, and Instagram still does not allow large-scale analysis of Reels. As such, our work is one of the few that analyze short-form videos and the first to analyze YouTube Shorts to the best of our knowledge. We now survey the existing works on short-form videos, which mostly rely on TikTok as data source.

Past analyses of TikTok mainly focused on its trending and recommendation algorithm. Klug et al. studied the characteristics of trending videos on TikTok and found that they have high video engagements and are more likely to be posted at certain times~\cite{klug2021trick}. They also report that using trending hashtags do not necessarily contribute to a video being trending. Simpson et al. investigated the impact of TikTok's algorithm on LGBTQ+ users and highlighted algorithmic exclusion and resilience in identity work~\cite{simpson2021you}. Boeker and Urman studied TikTok's recommendation algorithm and reported that it is influenced by users' location, language, and engagement with the content~\cite{boeker2022empirical}. Lee et al. employed a qualitative method in which they interview TikTok users to understand how they perceive and interact with TikTok's algorithms~\cite{lee2022algorithmic}. Other studies focused on how the platform shapes political communication by analyzing users' sentiment against the politicians~\cite{zeng2021okboomer}, hyperpartisan activity~\cite{medina2020dancing}, and the platforms' policies against misinformation~\cite{ling2023learn}.

\subsection{Impact of Platform Policies on Users}

Popular social media platforms introduce new policies from time to time. This may be due to external actors (e.g., governments) enforcing platforms to adopt new policies such as censorship~\cite{elmas2021dataset}, privacy policies~\cite{liu2021have}, and moderate hate speech~\cite{kozyreva2023resolving} and disinformation~\cite{papakyriakopoulos2022impact}. In other cases, platforms themselves may introduce changes to the platforms to enhance user experience. In all cases, such changes encourage users to change their behavior and adapt, which eventually shapes public communication. Such behavioral change may manifest itself in user content. For instance, users tend to use more abbreviations and contracted forms when they are constrained by the length of their content, but they also tend to create content with better quality~\cite{gligoric2018constraints}. They may also game platforms' policies to manipulate social media, such as purchasing popular accounts instead of growing new accounts~\cite{elmas2023misleading} or maintaining backup accounts to recover from platform suspensions~\cite{merhi2023information}. Other actors such as researchers may also be affected by platforms' data collection policies such as denying access to removed content~\cite{elmas2023impact} or limiting access to the API~\cite{braun2023journalism}. In this work, we study the impact of the introduction of Shorts on channel behavior by comparing the channels' regular videos and Shorts, which have not been studied to date.

\section{Conclusion}\label{sec:conclusion}

In conclusion, this article sheds light on the substantial impact of Shorts on the YouTube platform. We showed that channels which took an interest in Shorts, and uploaded at least one Short,  largely adopted the format, eventually surpassing RV uploads. 
We also showed that, while the collective amount of RV uploads stayed mostly constant over time, when looking at individual behaviors, most channels reduced their production of RVs, while increasing and then maintaining a constant frequency of Short uploads. 

We also observed that Shorts and RVs are not evenly distributed between content categories. Shorts mainly belong to lighthearted, entertainment categories, while RVs touch more diverse content, including news, politics, and education. 
This disparity in content production is reflected in content consumption. Indeed, the supremacy of Shorts in terms of views is less striking for entertainment-related categories than for the others. In art-related categories, Shorts barely attracted more views than RVs. 

Finally, we showed that Shorts progressively generated more views per video until getting five times more views on average than RVs, by the end of 2022. When looking at videos from the same channel, Shorts generated 110 times more views than their RVs counterparts. This effect is more pronounced for popular channels than for the rest.

\paragraph{Limitations \& Future Work}

We acknowledge that our findings are not free from limitations. As mentioned in the data collection, our dataset is biased towards the keywords we selected to collect our data. Additionally, we focused on channels who took an interest in Shorts and posted at least one Short. Opening the lens of focus to all types of channels to analyze the prevalence of channels that adopted Shorts, and overall channels behavior would contribute to ground this work into the current YouTube landscape. As such, further investigations comparing channels that adopted Shorts, and channels that did not, would provide interesting and complementary insights to this study. 
During our analysis of the engagement metrics we did not use comments' text. By collecting comments, we could observe the adaptation of the (commenting) user-base by analyzing the evolution of the network of commenters of a channel and the nature of the comments themselves. 


\begin{acks}
We thank Kévin Huguenin for his valuable insights and feedback on an earlier version of this paper. 
\end{acks}

\bibliographystyle{ACM-Reference-Format}
\bibliography{aaai22}

\end{document}